\documentclass[prd,twocolumn,showpacs,preprintnumbers,superscriptaddress,floatfix,nofootinbib]{revtex4-1}

\usepackage{multirow}

  \usepackage{float}
\usepackage{amsfonts}
\usepackage{graphicx,,booktabs}
\usepackage{amssymb}
\usepackage{amsmath,txfonts,ulem}
\usepackage{graphicx}
\usepackage{dcolumn}
\usepackage{color}
\usepackage{bm}
\usepackage{ulem}
\usepackage[colorlinks,
      citecolor=blue,
      anchorcolor=green,
      menucolor=orange,
      linkcolor=blue,
      filecolor=red,
      runcolor=pink,
      urlcolor=blue,
      frenchlinks=red]{hyperref}

\allowdisplaybreaks[4]

\begin{document}

\title{\boldmath Gluon gravitational form factors of   protons from   charmonium photoproduction }

\author{Xiao-Yun Wang}
\email{xywang@lut.edu.cn}
\affiliation{Department of physics, Lanzhou University of Technology,
Lanzhou 730050, China}
\affiliation{Lanzhou Center for Theoretical Physics, Key Laboratory of Theoretical Physics of Gansu Province, Lanzhou University, Lanzhou, Gansu 730000, China}

\author{Fancong Zeng}
\email{ zengfc@ihep.ac.cn}
\affiliation{ Institute of High Energy Physics, Chinese Academy of Sciences, Beijing 100049, China}
\affiliation{ University of Chinese Academy of Sciences, Beijing 100049, China}

\author{Quanjin Wang}
\affiliation{Department of physics, Lanzhou University of Technology,
Lanzhou 730050, China}

\date{\today}
\begin{abstract}

Inspired by the recent near-threshold $J/\psi$  photoproduction measurements, we discuss gluon gravitational form factors (GFFs) and internal properties of the proton. This work presents a complete analysis of the proton gluon GFFs connecting the gluon part of the energy-momentum tensor and the heavy quarkonium photoproduction. In particular, a global fitting of the $J/\psi$  differential and total cross section experimental data is used to determine the gluon GFFs as functions of the squared momentum transfer $t$. Combined with the quark contributions to the $D$-term form
factor extracted from the deeply virtual Compton scattering experiment, the total $D$-term is obtained to investigate
their applications in describing the proton mechanical properties. These studies provide a unique perspective on investigating the proton gluon GFFs and important information for enhancing QCD constraints on the gluon GFFs.

\end{abstract}

\maketitle

\section{introduction}

The  form  factor  provides  critical  information  about
many fundamental aspects of hadron structure. While the
weak and electromagnetic form factors of the proton are
well-established, our understanding on gravitational form
factors (GFFs) is incomplete. The GFFs are defined from
matrix elements of the quantum chromodynamics (QCD)
energy-momentum tensor  (EMT)  and  provide  direct   access  to  the  internal  structure  of  the  proton,  including  its
mass,  spin,  and  mechanical  properties \cite{Polyakov:2002yz,Polyakov:2018zvc}.
The  sum
contributions of  the  quark  and  gluon  GFFs  are   measurable  quantities  defined  purely  from  the  internal  system,
which describes the internal dynamics of the proton system \cite{Shanahan:2018nnv}.

The  GFFs,  including  the  $D$-form factor,  have  also
been  studied  in  numerous  frameworks
 	Recently,  the quark $ D$-form factor $D_q(t)$ has been extracted from the deeply virtual Compton scattering (DVCS) experiments  at the
Thomas Jefferson National Accelerator Facility (JLab), and the pressure distribution inside the proton has been reported \cite{Burkert:2018bqq}.
However, because the DVCS is almost
insensitive to gluons,
 the gluon $D$-form factor is  controversial and seldom extracted. On the theoretical side, one
study  obtained  the  proton  GFFs  and  investigated  the
mechanical  properties  using  a  light-front  quark-diquark
model  constructed  by  the  soft-wall  AdS/QCD   \cite{Chakrabarti:2020kdc}. The nucleon form factors of the  EMT  are studied in the framework of the Skyrme model and in-medium modified Skyrme model \cite{Cebulla:2007ei,Kim:2012ts}.
Reference \cite{Azizi:2019ytx} demonstrated the pressure, energy density, and mechanical radius of the nucleon in light-cone QCD sum rule formalism.
On the other hand, the distributions of pressure and shear forces inside the proton are investigated with lattice QCD  calculations  \cite{Shanahan:2018nnv,Shanahan:2018pib,Pefkou:2021fni}, enhancing our understanding of the proton GFFs.

Unfortunately,  there  are  no  experimental  constraints
on  the  gluon  GFFs  directly,  and  little  information  about
the gluon GFFs is explicit at present. However, at finite
momentum transfer,  the  near-threshold  heavy   quarkonium photoproduction, such as $J/\psi$ and $\Upsilon$ meson, offers a
superior  path  to  access  the  gluon  GFFs  \cite{Kharzeev:2021qkd,Hatta:2019lxo,Hatta:2018ina,Mamo:2019mka,Mamo:2022eui,Guo:2021ibg,Duran:2022xag,Ji:2020bby,Sun:2021gmi,Guo:2021ibg}. These processes have gained quite an interest in recent years because they promise to measure the naturalness of proton
mass decomposition \cite{Ji:2021mtz,Kharzeev:1998bz,Hatta:2018ina,Hatta:2019lxo,Guo:2021ibg}.
One reason is that the scalar gluon  operator  is  dominant  in  the  production   amplitude  of  heavy  quarkonium.  
Moreover,  heavy  vector
mesons photoproduction was employed because the high  mass of $J/\psi$ limits  the
interaction to a short distance. 
In electroproduction the short distance is given when $Q^2 >> 1\text{ GeV}^2$, i.e. high photon virtuality.
These
facts allow us to discuss the gluon GFFs by studying the
near-threshold photoproduction  data  of  heavy   quarkonium. Conversely, the connection between heavy quarkonium photoproduction  and  gluon  GFFs  also  faces   challenges. One study revealed that this process is light-cone
dominated and has no direct connection with gluon GFFs  \cite{Sun:2021pyw}.
Thereby,  more  theoretical  research  on  the  related
physical mechanisms is still needed.

Therefore, experimental information on vector meson
photoproduction is essential to gain insight into the gluon
GFFs  of  the  proton.  Recently,  the  GlueX  Collaboration
reported  the  near-threshold  cross  section  of  the  reaction  $\gamma p\to J/\psi p$ \cite{GlueX:2019mkq}.
The  JLab  experiment measured the  differential cross section of  $J/\psi$ on proton targets at a photon energy $E_{\gamma} $ from 9.1 GeV to 10.6 GeV \cite{Duran:2022xag}, which is a near-threshold energy region. Those experimental data offer  a good window for studying the internal character of the proton.
 Currently,  there are plans for future
experiments at JLab and Electron Ion Colliders (EICs) to
probe the deepest structure inside the proton and collect $J/\psi$ data \cite{Arrington:2021alx,Accardi:2012qut,Anderle:2021wcy}. High-precision experimental measurements are suggested to be performed at these facilities.

This  paper  is  organized  as  follows.  In  Sec.   \ref{sec:formalism},  we
provide  a  formulas  for  connecting  the  photoproduction
and gluon GFFs, and the process of calculating the proton  internal  properties.  In  Sec.  \ref{sec:result},  we  determine  the
gluon  GFFs  by  global  fitting  the  differential  and  total
cross section of  $J/\psi$ photoproduction. Subsequently, the
computation result  of  the  mechanical  properties  and   energy property inside the proton is presented. A summary
is given in Sec. \ref{sec:summary}.

\section{formalism}\label{sec:formalism}

\subsection{the photoproduction and gluon GFFs }
The $(00)$ component of the EMT defines the isotropous  form factor  in the Breit frame, which can be expressed as follows. \cite{Ji:2021pys,Ji:2021mtz,Kharzeev:1995ij,Polyakov:2018zvc}
\begin{align}\label{eq:M}
\left< P^{\prime}|T_{00}|P \right>=\bar{u} (P^{\prime}) u(P)G(t)
\end{align}
where the spinor normalization $\bar{u} (P^{\prime}) u(P)=2M$ and $M$ is the proton mass.  $G(t)$ is the proton GFFs, which are parametrized as follows. \cite{Polyakov:2018zvc,Ji:2021mtz}
\begin{equation}\label{eq:GT}
G(t)= M A_{q+g}(t) + \frac{t}{4M} B_{q+g}(t) - \frac{ t}{4 M } D_{q+g}(t)
\end{equation}
where    $t=-Q^2$ is the squared momentum transfer, the form factor  $B_{q+g}(t) =2J_{q+g}(t)-A_{q+g}(t)$ is consistent with zero   basically \cite{Mamo:2019mka,Mamo:2021krl,Shanahan:2018pib}. The form factors $A_{q+g}(t)$, $J_{q+g}(t) $ and $D_{q+g}(t)$ provide the information about the mass,  spin and the mechanical properties of the proton, respectively.

The  proton GFFs are the sum  contributions of the quark and gluon GFFs.Additionally,   the component of the gluon GFFs part can be written as follows. \cite{Polyakov:2018zvc}
\begin{equation}
 \begin{aligned}
G_g(t) = &    M A_{ g}(t)- \frac{ t}{4 M }\left( -B_{ g}(t)+D_g(t) \right)  +M\bar{C}_g(t)        \\
=&  \frac{3}{4}M A_{ g}(t) - \frac{ t}{4 M } D_{ g}(t) + \frac{3t}{16M} \left( B_g(t)+D_g(t)\right)
\end{aligned}
\end{equation}
because the $\bar{C}_g(t) $ form factor can be written as \cite{Tong:2022zax}
 $$\bar{C}_g(t)=-\frac{A_g(t)}{4} + \frac{- t}{16 M^2 }\left( B_{ g}(t)-3 D_{ g}(t)  \right) $$
 here, the constraint  $\bar{C}_g(t) +\bar{C}_q(t) =0 $ is due to EMT conservation  \cite{Polyakov:2018zvc}.

 Many   estimatations or models, including QCD sum
rule  \cite{Braun:2000kw,Braun:2006hz}, Braun-Lenz-Wittman  model, \cite{Braun:2006hz} and the asymptotic  model \cite{Tong:2022zax}, show  that the $B_g(t)+D_g(t)$  are order of magnitudes smaller compared to the $B_g(t)$ or $D_g(t)$  results \cite{Tong:2022zax}.
 In this paper, we mainly attribute the gluon  GFFs to the   first two terms in Eq. \ref{eq:G00}, as   the contribution of the $B_g(t)+D_g(t)$ is  negligible.  Therefore, the gluon GFFs are obtained as follows: 
\begin{equation}\label{eq:G00}
G_g(t)  \approx  \frac{3}{4} M A_{ g}(t) - \frac{ t}{4 M } D_{ g}(t)
\end{equation}

Next, we   demonstrate the complete   analysis of the proton gluon GFFs   connecting the gluon part of the EMT and the near-threshold  $J/\psi$   cross section.
Typically, the differential cross section of the $J/\psi$ photoproduction is given by \cite{ParticleDataGroup:2020ssz}
\begin{align}\label{eq:differential}
\frac{d\sigma_{\gamma p\to J/\psi p}}{dt}=\frac{1}{64 \pi W^2} \frac{1}{|{\bf p}_{\gamma }|^2} \left| \mathcal{M}_{\gamma p\to J/\psi p} \right|^2
\end{align}
where $W$ is the center of mass (c.m.)  energy and ${\bf p}_{\gamma }$ is the  c.m. photon momentum in the $ \gamma p\to J/\psi p$ process.
 As an assumption, the amplitude primarily attributes to the  gluon part of the EMT of QCD in this paper,  which can be written as follows. \cite{Kharzeev:2021qkd}
\begin{align}\label{eq:amplitude}
 \mathcal{M}_{\gamma p\to J/\psi p} =- 2 c_2 Q _c  M \left< P^{\prime}|g^2 T^g_{00}|P \right>
\end{align}
where $Q_e=2e/3$ represents the coupling of the photon to
the electric charge of the quarks in $J/\psi$ meson;
 $ c_2 $, the short-distance coefficient, is on the order of $\pi r_{c\bar{c}}^2$; and
$g^2=4$ is the QCD coupling with $\alpha_s \approx 0.32 $ \cite{Hatta:2018ina,Novikov:1980fa}.

By integrating the differential cross section (Eq. \ref{eq:differential})
over the allowed kinematical range from $t_{min}$ to $ t_{max}$,
  the total cross section are computed and  can be written as follows. 
\begin{align}\label{eq:total}
\sigma=\int_{t_{min}}^{t_{max}}  dt  \left(\frac{d \sigma}{d t} \right) ,
\end{align}
 here, the limiting values $t_{min}$ and  $t_{max}$ are
\begin{align}
t_{max}( t_{min})=  m_{J/\psi}^2-2E_{\gamma} E_{J/\psi}  \pm   2 |{\bf p}_{\gamma }||{\bf p}_{J/\psi}|  .
\end{align}
The energies and momenta of the  photon and vector meson in the  c.m.  frame are
\begin{equation}
\begin{aligned}
|{\bf p}_{J/\psi}|= \frac{1}{2W} \sqrt{(W^2-M^2)^2+ m_{J/\psi}^2(m_{J/\psi}^2-2W^2-2M^2)  }, \\
 E_{J/\psi}=\sqrt{{\bf| p}_{J/\psi}| ^2+m_{J/\psi}^2},~~ |{\bf p}_{\gamma }|= \frac{1}{2W} (W^2-M^2), ~~ E_{\gamma }= {\bf |p_\gamma |}.
\end{aligned}
\end{equation}
Thus  far, we have established the relationship between the proton GFFs and  $J/\psi$ photoproduction, including  the differential  and  total  cross section in Eqs. \ref{eq:differential} and \ref{eq:total}.

For  $ A_{q+g}(t)$, the mass distribution of the proton is encoded in the $A$-form factor, which can be expressed under the dipole form parametrization as  follows. 
\begin{align}
A_{q+g}(t)=\frac{A_{q}(0)}{(1- t/m_q^{ 2} )^2}+\frac{A_{g }(0)}{(1- t/m_g^2 )^2}
\end{align}
where the constraint $A_{q}(0)+A_{g}(0)=1$ is the consequence of momentum conservation \cite{Ji:1997pf,Polyakov:2018zvc}.
Moreover, the gluon contribution $A_g(0)=0.414$ was obtained from CT18 global QCD analysis \cite{Hou:2019efy},
 and   agrees  with  other LQCD results \cite{Pefkou:2021fni,Alexandrou:2020sml,Yang:2018nqn}. Therefore,  the parameter $A_g(0)$  in  $ A_g(t)$ is fixed in this study, and  $m_g$ is a free parameter determined  by fitting  experimental data.

The $D$-form factor $D_{q+g}(t)$ is  an area of significant interest, which has attracted considerable attention recently \cite{Polyakov:2018zvc,Duran:2022xag}.
 The gluon $D$-form factor $D_g(t)$ is typically  parameterized in the tripole form and provided as follows. \cite{Fiore:2021eav,Burkert:2018bqq}
 \begin{align}\label{eq:Ggluon}
	D_g(t)=\frac{D_g(0)}{(1-t/d_g^2)^{3}}
\end{align}
where
$D_g(0) $ and $d_g$ are free parameters adjusted to the experimental data. Note that $D_g(0) $ is negative as the  pressure distribution is found to be repulsive near the  proton center.

It has been determined that the form factor $G(t)$ in Eq. \ref{eq:GT} and \ref{eq:G00} at the momentum transfer $t$=0  satisfies
 \begin{align}
G(0)=M \text{~~~and~~~}
G_g(0)=\frac{3}{4}MA_g(0)
\end{align}
Therefore, the coefficient $ c_2 $ can be determined by extracting the near-threshold differential cross section at $t$=0, which can be written as follows. 
 \begin{align}\label{eq:dt=0}
\left.\frac{d\sigma_{\gamma p\to J/\psi p}}{dt}\right|_{t=0}=\frac{1}{64 \pi W^2} \frac{1}{|{\bf p}_{\gamma }|^2} \left|3 c_2 Q _c   g^2  M^3 A_g(0) \right|^2
\end{align}
As the  differential cross section at squared momentum transfer $t$=0 is nonphysical with no experimental measurement,
we will identify the left side of Eq. \ref{eq:dt=0} at different c.m. energy based on the  model prediction  as discussed in Ref.  \cite{Wang:2022vhr}, which is  described in detail  in Sec. \ref{sec:result}.  As  a  result,  reliable  gluon  GFFs  can  be  obtained while   avoiding the c.m. energy dependence caused by the differential cross sections at    varying    photon energies. This  approach  provides  significant  information  on  the
gluon GFFs.  

 Finally, we construct the   relationship between the gluon GFFs of the proton  and the  near-threshold  heavy quarkonium photoproduction. Thus one can derive the  gluon GFFs, which is the joint effect of  $J/\psi$ differential and total cross section.

\subsection{proton internal properties}

The pressure $p(r)$ and shear forces $s(r)$ are “good observables” to report the pressure and shear forces distributions, indicating that the average value of the directional static pressure and shear forces along the three Cartesian axes can be expressed as follows.  \cite{Polyakov:2018zvc,Polyakov:2002yz}
\begin{align}\label{eq:sr}
 s_{q+g}(r) = - \frac{1}{2} r\frac{d }{dr} \frac{1 }{ r} \frac{d }{dr} \widetilde{D} (r)
\end{align}
\begin{align}\label{eq:pr}
  p_{q+g}(r) = \frac{1}{3} \frac{1 }{ r^2} \frac{d }{dr} r^2 \frac{d }{dr} \widetilde{D} (r)
\end{align}
Here $\widetilde{D} (r)$ is the Fourier transform of $D_{q+g}(t)$ and can be
expressed as follows.   \cite{Polyakov:2018zvc,Polyakov:2002yz}
\begin{equation}
\begin{aligned}
 \widetilde{D} (r)&=\int \frac{d^3 {\bf \Delta} }{2 M(2\pi)^3 } e^{-i {\bf \Delta} r} D_{q+g}(- {\bf \Delta}^2)
 \\
&=\int \frac{d^3 {\bf \Delta} }{2 M (2\pi)^3 } e^{-i {\bf\Delta} r}\left( D_{g}(- {\bf\Delta}^2) +D_{q}(- {\bf\Delta}^2) \right)\\
\end{aligned}
\end{equation}
 Note that the pressure distribution $ r^2 p_{q+g}(r)$ satisfies the internal forces balance inside a composed particle based on \cite{Polyakov:2018zvc}
\begin{align}\label{eq:r2pr}
\int_0^{\infty}  dr ~r^2 p_{q+g}(r) =0
\end{align}
The normal   forces in the composed particle system can be written as  \cite{Polyakov:2018zvc}
\begin{align}
 F_n(r)=\frac{2}{3} s_{q+g}(r)+p_{q+g}(r)
 \end{align}
 here the positive and negative eigenvalues correspond to
“stretching" or “squeezing" along the corresponding principal axes, respectively.
The normal  forces satisfy $F_n(r)>0$ \cite{Polyakov:2018zvc}.
One can define the proton mechanical radius in terms of the normal   forces in the proton, which can be written as \cite{Polyakov:2018zvc}
\begin{equation}\label{eq:mech}
\left<R^2_{mech}\right>  = \frac{\int d^3r ~r^2  F_n^{ q+ g }(r)    }{\int d^3r ~ F_n^{q+ g  }(r)    }   = \frac{ 12 \left( \frac{ D_q(0) }{ d_q}+ \frac{D_g(0)}{ d_g} \right) }{    D_q(0) d_q    +  D_g(0) d_g  }
\end{equation}
After calculating the form factor   $D_{q+g}(t)$, the   pressure in the proton center can be
computed directly as \cite{Polyakov:2018zvc}
\begin{equation}\label{eq:p0}
p(0)=\frac{1}{24 \pi^{2} M} \int_{-\infty}^{0}   - (-t)^{3/2}   D_{q+g}(t) d t
\end{equation}
which is consistent with the illustration in Eq.  \ref{eq:pr}.

The total energy density $T_{00} (r)$ satisfied $T_{00}(r)>0$ in a mechanical system is defined for the total system in Eqs. \ref{eq:M} and \ref{eq:GT}, which  can be written as \cite{Polyakov:2018zvc}
 \begin{equation}\label{eq:T00g} 
 \begin{aligned} 
&T_{00}(r)=  \int \frac{d^{3} {\bf\Delta} }{(2 \pi)^{3}} e^{-i r {\bf\Delta} } \left(   MA_{q+g}(t)-\frac{t}{4M}D_{q+g}(t)    \right) \\
&=\sum_{a=q,g}\frac{ 16 M^2      A_a(0)  m_a^3 e^{-m_a r}   - 
 (-3 + d_a r) D_a(0 ) d_a^5 e^{-d_a r} }{ 128 \pi M}
\end{aligned}
\end{equation}
where $-{\bf\Delta}^2= t$. 
The total energy density $T_{00}(r)$ satisfy the following condition: 
 \begin{equation}
 \begin{aligned}
\int d^{3} r T_{00}(r)&=\int d^{3} r \left(T^g_{00}(r)+ T^q_{00}(r)\right)\\
&= M \left(A_{q }(0)+A_{g }(0)\right)=M
\end{aligned}
\end{equation}

  The energy density satisfies $T_{00}(r)>0$ in a mechanical system,   allowing us to introduce the mean square radius of the energy density as follows. \cite{Polyakov:2018zvc}
 \begin{equation}
 \left\langle R_{E}^{2}\right\rangle =\frac{\int d^{3} r r^{2} T_{00}(r)}{\int d^{3} r T_{00}(r)} =6A_{q+g}^{\prime}(0)-\frac{3\left(D_0+d(0)\right)}{2M^2}
 \end{equation}

\section{RESULTS AND DISCUSSION}\label{sec:result}
\begin{figure}[H]
	\includegraphics[width=0.48\textwidth]{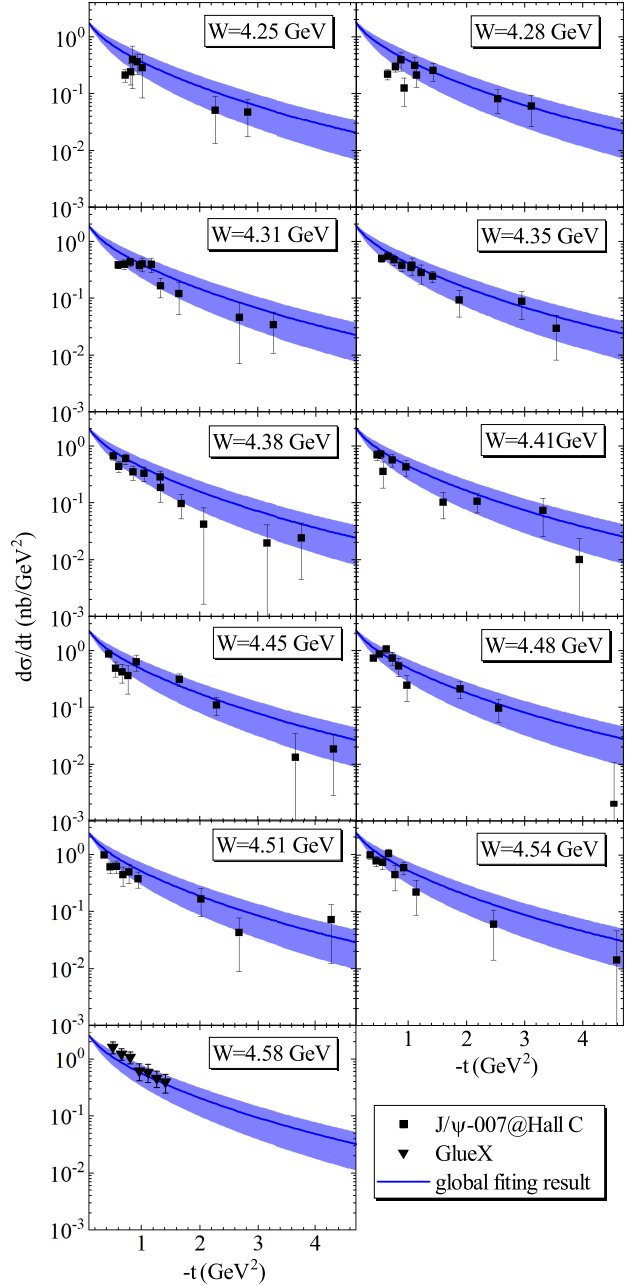}
\caption{ Global fitting result of $\gamma p \to J/\psi p$ differential cross section as a function of $-t$ at c.m. energy $W=4.25 \text{ GeV},$ $ 4.28 \text{ GeV}, $ $4.31 \text{ GeV}, $ $4.35 \text{ GeV},$ $ 4.38 \text{ GeV} ,$ $4.41 \text{ GeV}, $ $ 4.45 \text{ GeV}, $ $4.48 \text{ GeV},$ $ 4.51 \text{ GeV},$ $ 4.54 \text{ GeV}$ and $4.58 \text{ GeV}$. The blue bands reflect  statistical error of $ m_g$, $ d_g$ and $D_g(0)$. References of data can be found in \cite{Duran:2022xag,GlueX:2019mkq}. }
 \label{cross-section}
\end{figure}

\begin{figure}[htbp]
	\includegraphics[width=0.46\textwidth]{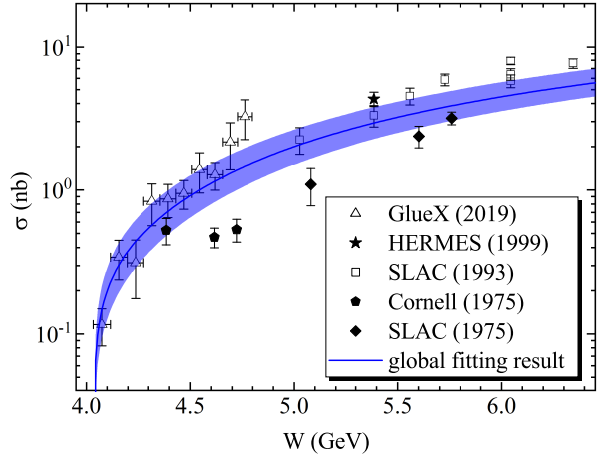}
\caption{ Global fitting result of $\gamma p \to J/\psi p$ total cross section as a function of   c.m. energy $W $. The blue band reflects  statistical error of $ m_g$, $ d_g$, and $D_g(0)$. References of data can be found in   \cite{Gittelman:1975ix,GlueX:2019mkq,Camerini:1975cy,Frabetti:1993ux,Amarian:1999pi} . }
 \label{total-cross-section}
\end{figure}

 In our previous work  \cite{Wang:2022vhr}, the analysis  revealed that certain light quarks are strongly suppressed in  heavy quarkonium photoproduction  and are likely to dominate  the two-gluon exchange mechanisms.
 The obtained
numerical results showed that the two-gluon exchange model can explain the near-threshold $J/\psi$ photoproduction  experimental data well \cite{Wang:2022vhr}. Consequently, 
the   differential cross section $d\sigma/dt|_{t=0} $ in Eq. \ref{eq:dt=0} at various c.m. energy is predicted by the two-gluon exchange model, allowing  the  identification  of  the  corresponding  short-distance coefficient $ c_2 $ can be identified. Additionally, the introduction of the model is helpful  for  obtaining  a  continuous  total cross section. Subsequently, the gluon GFFs in Eq. \ref{eq:G00}  are   achieved through fitting Eqs. \ref{eq:differential} and \ref{eq:total}  simultaneously.
 By global fitting the near-threshold $t$-dependence  $J/\psi$ differential cross section and $W$-dependence  total cross section experimental data \cite{Duran:2022xag,GlueX:2019mkq,Gittelman:1975ix,GlueX:2019mkq,Camerini:1975cy,Frabetti:1993ux,Amarian:1999pi}, the free parameters $m_g$, $ d_g$ and $D_g(0) $ in Eq. \ref{eq:G00}   are computed.
The differential and total experimental data used in this study   are derived from the experiment results  that are currently closest to the threshold.
The comparison between the $J/\psi$ photoproduction (blue solid curves)
  and   experimental measurements  (black points)
 are  presented in Figs. \ref{cross-section} and \ref{total-cross-section}, showing a good agreement. The blue bands reflect  a statistical error of parameters  $ m_g$, $ d_g$, and $D_g(0)$.  The results of holographic QCD and the GPD+VMD approach  were recalculated in Ref. \cite{Duran:2022xag}  using the latest differential cross section data.
 The obtained   gluon GFFs   are compared to that of the  holographic QCD, the GPD+VMD approach, and LQCD \cite{Shanahan:2018pib,Mamo:2022eui,Pefkou:2021fni,Guo:2021ibg,Duran:2022xag}, as presented  in Table \ref{tab:D}.
Notably, our results are comparable to that of holographic QCD determination and LQCD calculation.

\begin{figure}[htbp]
	\centering
	\includegraphics[width=0.45\textwidth]{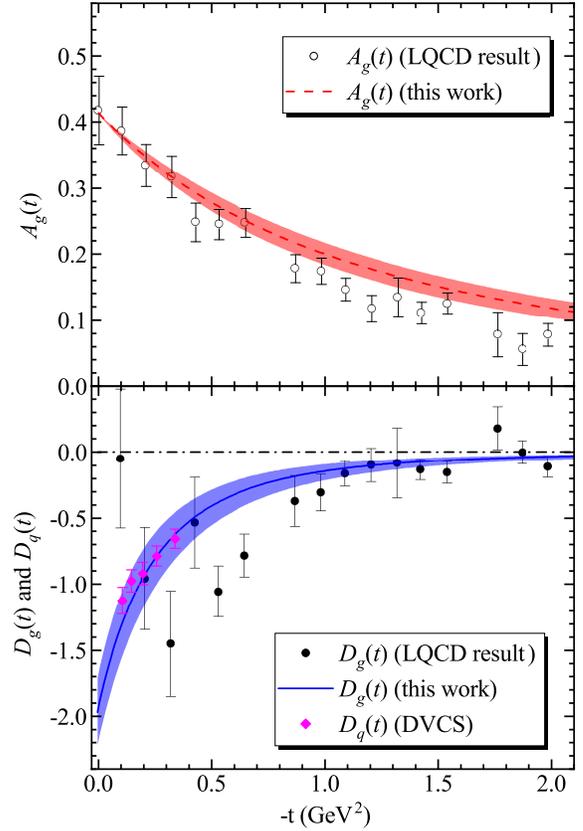}
\caption{ Top panel:  The  gluon $A$-form factor $A_{g}(t)$ (red-dashed  curve) compared with the LQCD determinations  \cite{Pefkou:2021fni}.
 Bottom panel:
The gluon $D$-form factor   $D_g(t)$ (blue-solid curve) compared with the LQCD determinations  \cite{Pefkou:2021fni} and the quark $D$-form factor from DVCS enperiment \cite{Burkert:2018bqq}.       The blue band reflects an statistical error of the parameters
  $ m_g$, $ D_g(0)$ and $d_g$.  }
 \label{DqDg}
\end{figure}

As shown in Fig. \ref{DqDg}, the values of gluon $A$-form factor $A_g(t)$ (red dashed curve) and  gluon $D$-form factor $D_g(t)$ (blue solid curve) are compared with the LQCD determinations  \cite{Pefkou:2021fni}. Here the error of parameters  $ m_g$, $ D_g(0)$ and $d_g$ include all uncertainties of $A_g(t)$ and $D_g(t)$.  One finds that the results obtained for gluon  $D$-form factor in this work is
comparable with that  obtained  from  the LQCD computations, while  the  values  of $A_g(t)$  is bigger than the LQCD results slightly.
We have also compared the gluon $D$-form factor with the quark counterparts  extracted  from the DVCS experiment, and as a result, the gluon and quark $D$-form factor are approximately  comparable, which is in agreement   with numerous previous studies \cite{Burkert:2018bqq,Pefkou:2021fni,Shanahan:2018pib}.

 The sum of the quark and gluon $D$-form factors  $D_{g+q}(t)$ is a measurable quantity defined solely by   the $D$-term inside the proton.
Particularly,  one  can  obtain  the   quark $D$-form factor by fitting the DVCS data \cite{Burkert:2018bqq} using the tripole form presumption \cite{Fiore:2021eav}. Combined with the gluon $D$-term achieved in this work, one can obtain the proton mechanical properties  from $D_{g+q}(t)$, including the quark and gluon contributions.

\begin{figure*}[htbp]
	\centering
	\includegraphics[width=0.43\textwidth]{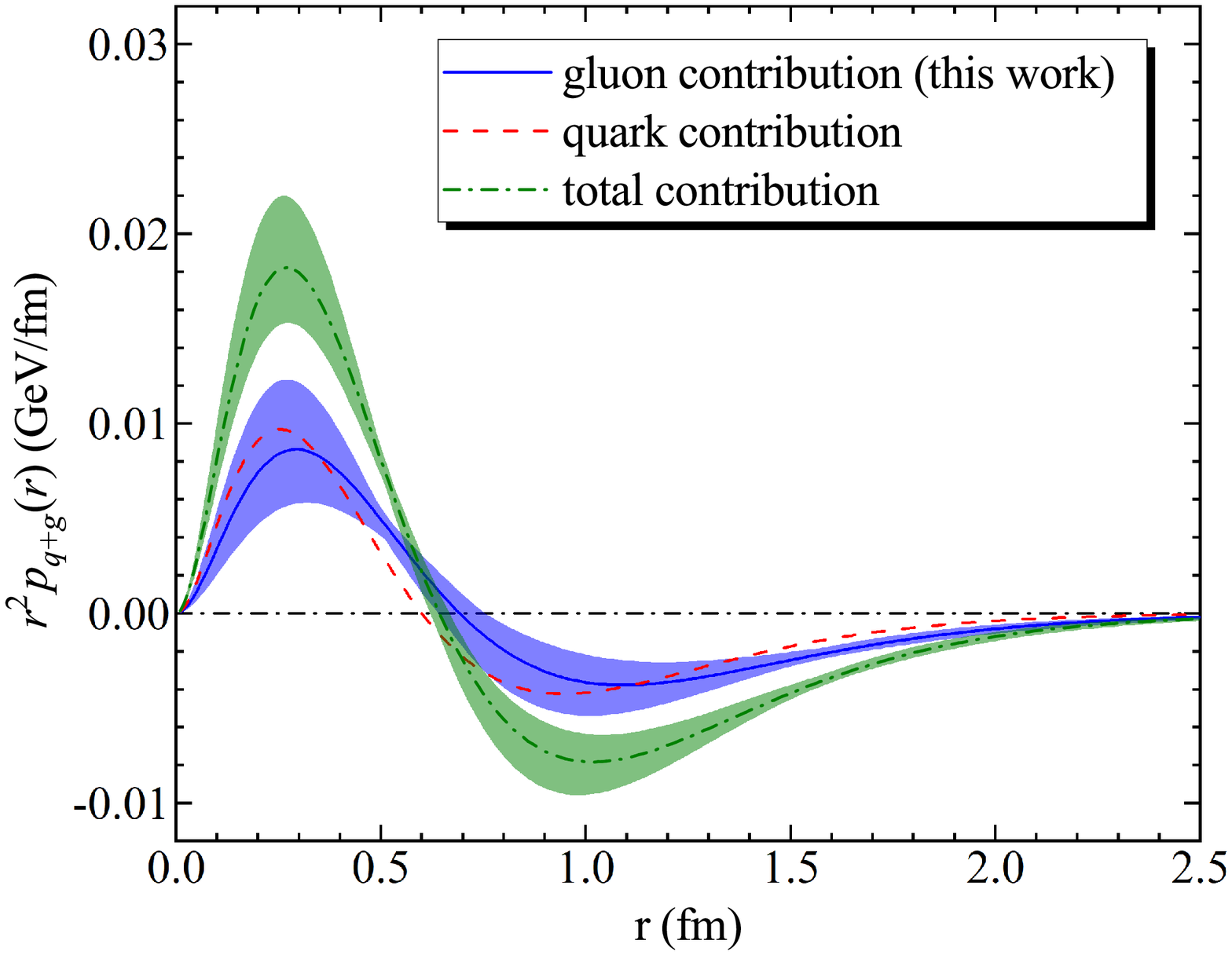}
	\includegraphics[width=0.43\textwidth]{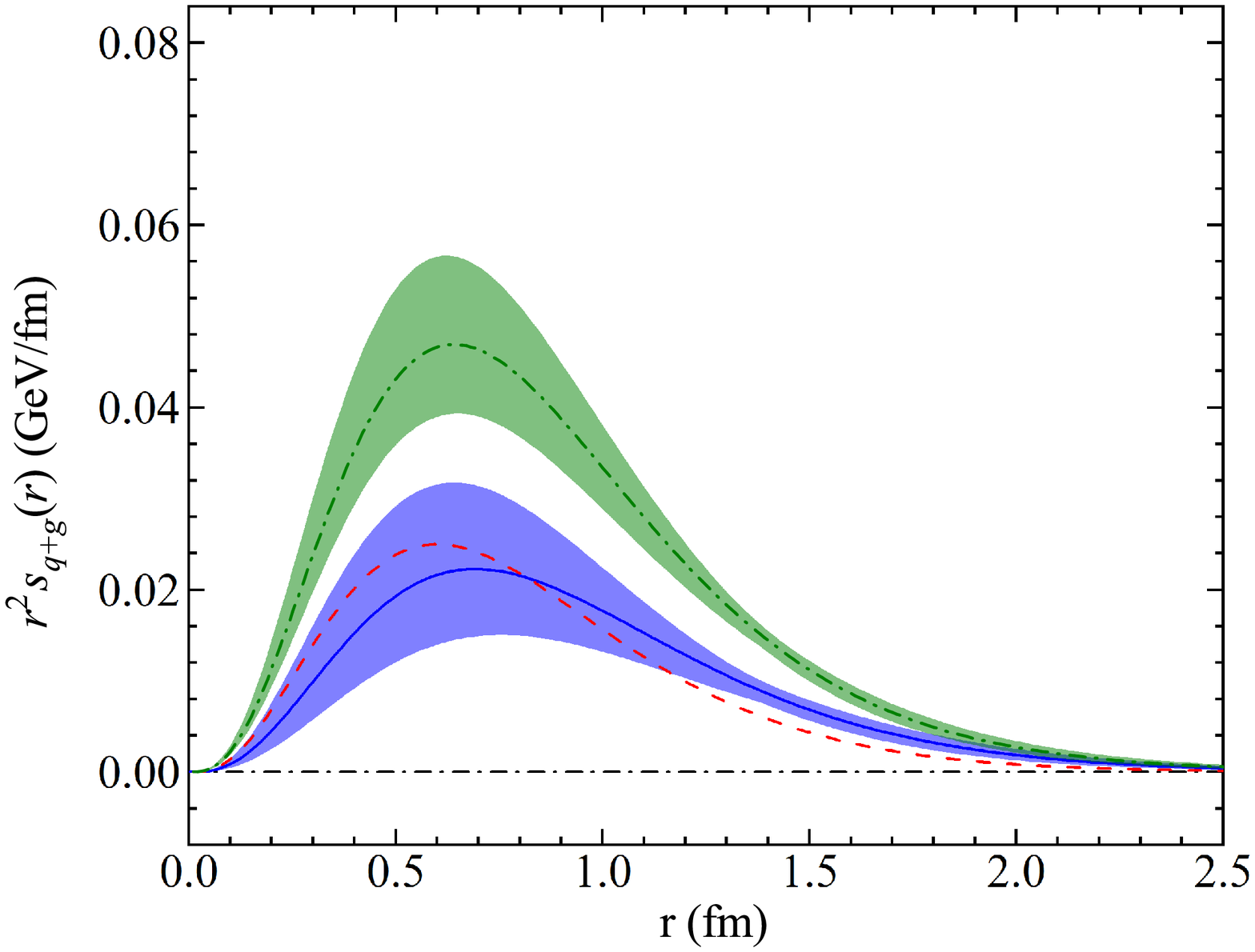}
\caption{  Left  panel:  The pressure distribution $r^2p(r)$    inside the proton. Right panel: The shear force  distribution $r^2s(r)$ inside the proton.   The red-dashed and blue-solid curve shows the  quark and gluon contributions,   respectively. The green-dot-dashed curve represents  the total pressure and shear forces distributions. The blue and green band cover all the statistical uncertainties of $D_g(0)$ and $ d_g$. }
 \label{r2pr}
\end{figure*}

The pressure and shear forces distributions inside the proton are achieved and displayed in Fig. \ref{r2pr}. The red-dashed and blue-solid curves indicate the gluon and quark contributions of the pressure and shear forces distributions, respectively. The blue and green bands represent the uncertainties that come from the error of the parameters $ d_g$  and $D_g(0)$. Here,the positive sign indicates repulsion   toward the outside, and the negative sign indicates attraction directed towards the inside.
The total pressure and shear forces contributions for the sum of the quark and gluon contributions are illustrated as the green-dot-dashed curve in Fig. \ref{r2pr}.
 It was found that the
pressure is positive in the inner region  and negative in the outer region,
with a zero crossing near $ r = 0.67 \text{ fm}$, which shows that the repulsive and binding pressures dominate in the proton and are separated in radial space.
Moreover, the shear forces distribution reaches its peak near $ r =  0.63$ fm in our observation.

 \begin{table}
\caption{  Parameters $ m_g$, $ D_g(0)$ and $d_g$ by a global fitting of the differential and total cross section experimental data \cite{Duran:2022xag,GlueX:2019mkq,Gittelman:1975ix,GlueX:2019mkq,Camerini:1975cy,Frabetti:1993ux,Amarian:1999pi}, compared with the holographic QCD, the GPD+VMD approach and LQCD results \cite{Shanahan:2018pib,Mamo:2022eui,Pefkou:2021fni,Guo:2021ibg,Duran:2022xag}. }
\begin{tabular}{ccccccc }
\hline\hline\noalign{\smallskip}
 approach & $m_g$ (GeV) & $D_0$ & $d_g$ (GeV) \\ \noalign{\smallskip}\hline\noalign{\smallskip}
 Holographic QCD \cite{Mamo:2022eui,Duran:2022xag}  & \multirow{2}*{1.575} & \multirow{2}*{$-1.80\pm 0.528$} & \multirow{2}*{$1.21\pm 0.21$} \\
tripole-tripole &&&\\\noalign{\smallskip}\hline\noalign{\smallskip}
GPD + VMD \cite{Guo:2021ibg,Duran:2022xag} & \multirow{2}*{2.71} & \multirow{2}*{$-0.80\pm 0.44$} & \multirow{2}*{$1.28\pm 0.50$} \\
tripole-tripole &&&\\\noalign{\smallskip}\hline\noalign{\smallskip}
LQCD \cite{Shanahan:2018pib} & \multirow{2}*{1.641} & \multirow{2}*{$-1.932\pm 0.532$} & \multirow{2}*{$1.07\pm 0.12$}\\
 tripole-tripole &&&\\\noalign{\smallskip}\hline\noalign{\smallskip}
LQCD \cite{Pefkou:2021fni} & \multirow{2}*{1.13} & \multirow{2}*{$-10.0$} & \multirow{2}*{$0.48$}\\
dipole-dipole &&&\\\noalign{\smallskip}\hline\noalign{\smallskip}
this work & \multirow{2}*{$1.51\pm 0.10$} & \multirow{2}*{$-1.97 \pm 0.25$} &\multirow{2}*{ $0.86 \pm 0.07$} \\
dipole-tripole &&&\\
\noalign{\smallskip}
\hline\hline
\end{tabular}
\label{tab:D}
\end{table}

After discussing the gluon form factors $A_{g}(t)$ and $D_{g}(t)$,
 the gluonic contribution to the nucleon mechanical properties  can  be  achieved. 
An  important  mechanical quantity,  known as $p_g(0)$, denotes the pressure of the gluon contribution in the center of the nucleon and   has a of $ 0.62_{-0.27}^{+0.42} \text{ GeV/fm}^3 $. One can add the quark contribution $p_q(0)=0.93\text{ GeV/fm}^3$ and  compute  the  system pressure $p(0)=1.55_{-0.27}^{+0.42}  \text{ GeV/fm}^3 $ at the center of the proton.
Moreover, the proton mechanical radius is computed to be $0.75_{-0.03}^{+0.04} \text{ fm}$.
As listed in Table \ref{tab:approach}, those proton mechanical quantities are compared with other existing theoretical results. As shown   in Table \ref{tab:approach}, our statements on the pressure  density differ from the findings in previous studies considerably \cite{Azizi:2019ytx,Azizi:2019ytx,Anikin:2019kwi,Choudhary:2022den,Jung:2014jja,Kim:2012ts}. In fact, the quark contribution to the pressure is bigger than most of those reported previously, regardless of the gluon contribution.
 The calculation of mechanical radius  is consistent with the result reporetd in Refs. \cite{Azizi:2019ytx,Anikin:2019kwi,Anikin:2019kwi,Shanahan:2018nnv,Jung:2014jja}, subjected to the error margin.

 \begin{table}
\centering
\caption{Numerical values of mechanical quantities of proton, including the proton mechanical radius and the pressure at the center of the proton, compared with other predictions from different approaches. }
\setlength{\tabcolsep}{-0.1mm}{
\begin{tabular}{ccccc}
\hline\hline\noalign{\smallskip}
  Approaches and Models ~~  &   $p(0)$ (GeV/fm$^3$)     & $ \sqrt{\left<R^2_{mech}\right>}  $ (fm)   \\
\noalign{\smallskip}\hline\noalign{\smallskip}
 light-front quark-diquark model  \cite{Choudhary:2022den}   & 4.76 &   0.50  \\
\noalign{\smallskip}\hline\noalign{\smallskip}
Skyrme model \cite{Kim:2012ts}~~~~~ & 0.26  & - \\
\noalign{\smallskip}\hline\noalign{\smallskip}
light-cone QCD \cite{Azizi:2019ytx} ~~~~~& 0.67   & 0.73
  \\
\noalign{\smallskip}\hline\noalign{\smallskip}
light-cone sum rules at leading order \cite{Anikin:2019kwi} &0.84  & 0.72  \\
\noalign{\smallskip}\hline\noalign{\smallskip}
lattice QCD (modified $z$-expansion) \cite{Shanahan:2018nnv} & -  & 0.71 \\
\noalign{\smallskip}\hline\noalign{\smallskip}
lattice QCD (tripole ansatz) \cite{Shanahan:2018nnv} &-   &0.75   \\
\noalign{\smallskip}\hline\noalign{\smallskip}
$\pi-\rho-\omega$ soliton model \cite{Jung:2014jja} & 0.58   & -   \\
\noalign{\smallskip}\hline\noalign{\smallskip}
this work ~~~~~ & $1.55_{-0.27}^{+0.42} $   &  $0.75_{-0.03}^{+0.04} $   \\
\noalign{\smallskip}
\hline\hline
\end{tabular}}
\label{tab:approach}
\end{table}

\section{summary}\label{sec:summary}
This work constructs a connection between the gluon part of EMT and the near-threshold    charmonium photoproduction.  The gluon GFFs,  as functions of the squared momentum transfer $t $,  are determined by a global fitting of the  $J/\psi$ differential and total cross section experimental data. 
All  gluon form factors $A_g(t),B_g(t)$, $D_g(t)$  and $\bar{C}_g(t)  $, which are related to different components of the gluon GFFs, are resolved.
One finds that $D_g(t)$ is comparable with the lattice QCD results, while  the value of $A_g(t)$  is slightly bigger than the   holographic QCD and LQCD results. Subsequently,
the gluon contribution of the energy density in the center of the proton, which  are  determined by both $A_g(t)$ and $D_g(t)$,  is  calculated.
Combined with the quark $ D$-form factor extracted from the DVCS experiment, the total $ D$-term $D_{q+g}(t)$ can be used to investigate its  potential  applications  in  describing   the mechanical properties. Consequently, the pressure and shear forces distributions inside the proton,  including the gluon and quark contributions, are obtained.
 
It has been suggested that the value of the proton charge radius is  \cite{ParticleDataGroup:2020ssz}
\begin{equation*}
R_C=0.8409 \text{ fm}
\end{equation*}
The proton mechanical radius  we obtained   is estimated to be  $0.75_{-0.03}^{+0.04} $ fm,   which is slightly less than the charge radius.
Generally, the measurements of the charge distribution  and  the  mechanical  properties  of  the  proton  can
contribute to our understanding of the origin of the proton structure.  This  study  provides  useful  theoretical   insights for the QCD constraints on the gluon GFFs of the
proton.

In fact, the dipole and tripole form are typically considered in the $A_{q+g}(t)$ and $D_{q+g}(t)$   form factors,  allowing for feasible  fitting results in this study. Moreover, this ansatz of the gluon GFFs is convenient to compare with the quark GFFs and other theoretical studies. Nevertheless,  it may be feasible to achieve global fitting using artificial neural networks or  Schlessinger
Point Method \cite{Dutrieux:2021nlz,Cui:2022fyr,Cui:2021vgm}.  Additionally, these new approaches can   investigate  the model outcomes  for  $d\sigma/dt|_{t=0} $.
Therefore, this work is only the first step, and we will  optimize the computational methods in  future  studies.

The high-precision photo/electroproduction data of vector mesons serve as a crucial foundation  for the   accurate study of the internal structural properties of the proton. As  a  result,  we  recommend    relevant  experimental measurements based  on  our  findings  to  be  conducted   at JLab \cite{GlueX:2019mkq} or EICs \cite{Accardi:2012qut,Anderle:2021wcy} facilities.

\section{Acknowledgments}

This work is supported by the National Natural Science Foundation of China under Grant Nos. 12065014 and 12047501, and by the Natural Science Foundation of Gansu province under Grant No. 22JR5RA266. We acknowledge the West Light Foundation of The Chinese Academy of Sciences, Grant No. 21JR7RA201.

\end{document}